\documentclass[aps,twocolumn,groupedaddress,showpacs]{revtex4}
\usepackage{graphicx}%

\begin{document}
\bibliographystyle{apsrev}


\title{Ten-nanometer surface intrusions in room temperature silicon}


\author{Shuhan Lin}
\author{Iris Mack}
\author{Noom Pongkrapan}
\author{P. Fraundorf}
\email[]{pfraundorf@umsl.edu}
\affiliation{Physics \& Astronomy and Center for Molecular Electronics, 
U. Missouri-StL (63121) \\ St. Louis, MO, USA}


\date{\today}

\begin{abstract}
  Defects $\sim 10 nm$ in size, with number densities $\sim 10^{10} cm^{-2}$, form spontaneously beneath ion-milled, etched, or HF-dipped silicon surfaces examined in our Ti-ion getter-pumped transmission electron microscope (TEM) after exposure to air.  They appear as weakly-strained non-crystalline intrusions into silicon bulk, that show up best in the TEM under conditions of strong edge or bend contrast.  If ambient air exposure is $< 10$ minutes, defect nucleation and growth can be monitored {\em in situ}.  Possible mechanisms of formation are discussed.
\end{abstract}
\pacs{61.72.Ff, 68.37.Lp, 81.05.Cy, 85.40.-e}

\maketitle

\section{Introduction}

As gigascale integrated circuit devices approach nanometer dimensions, 
the bulk {\em and} surface structure of single crystal silicon is 
attracting finer scrutiny.  In this context, recent searches for bulk 
defects 10 nm in size required that we consider more carefully some 
ubiquitous but weakly-strained specimen preparation artifacts that 
for studies of larger bulk defects we had ignored.

TEM observations show that these imperfections, which form in variously-oriented silicon surfaces after exposure to air, are strikingly similar after thinning with physically different methods.  Moreover, the process of their formation can be slowed to laboratory time scales at room temperature if air exposure is sufficiently short.  This allows {\em in situ} study of nucleation and growth, offering a view of dynamic processes in silicon which take place on laboratory time scales at room temperature, including the movement of self-interstitials and surface diffusion barrier formation.  Both of these are subjects of widespread fundamental and applied interest \cite{Aziz97, Newman00, Myers00, Estreicher01}. 

\section{Experimental Setup}

This study was begun on 200mm p-type boron-doped wafers, although the results have been checked on archive specimens with a wide range of ingot diameters (down to 100mm), growth conditions, impurity concentrations, and thermal histories.  The silicon was cut into 3 mm diameter disks, mechanically thinned, and dimpled to about 20 micron thickness in the center.  Final thinning to perforation involved one of three methods:  (a) ion-milling, in which the thinned disk was bombarded with 1 keV or 4 keV Ar ions at an incidence angle of 70 to 85 degrees, until perforation; (b) chemical etching, in which the thinned disk was put into a solution of concentrated $HNO_{3} : HF = 3:1$ until perforation, followed by a water wash; and (c) ``HF dip", in which a chemically etched specimen is dipped for 30 seconds in concentrated 50-55$\%$ HF before the wash, so as to passivate surface silicon atoms with hydrogen to reduce the rate of surface oxidation.  

Specimens were examined in a 300 kV Philips EM430ST TEM with point resolution near $0.2 nm$.  A $10 \mu m$ diameter objective aperture was used to increase diffraction contrast, as an aid in locating otherwise nearly invisible defects.  Choice of regions and orientations where strong thickness fringe and/or bend contour contrast was available \cite{Hirsch65} also aided visualization.  This sometimes involved enlarging the perforation to increase "wedge angles" at the perforation edge, since contrast in the neigborhood of thickness fringes \cite{AshbyBrown63b} was the most reliable way to count number densities over large areas.

\section{Observations of defect structure and abundance}

\begin{figure}[tbp]
\includegraphics{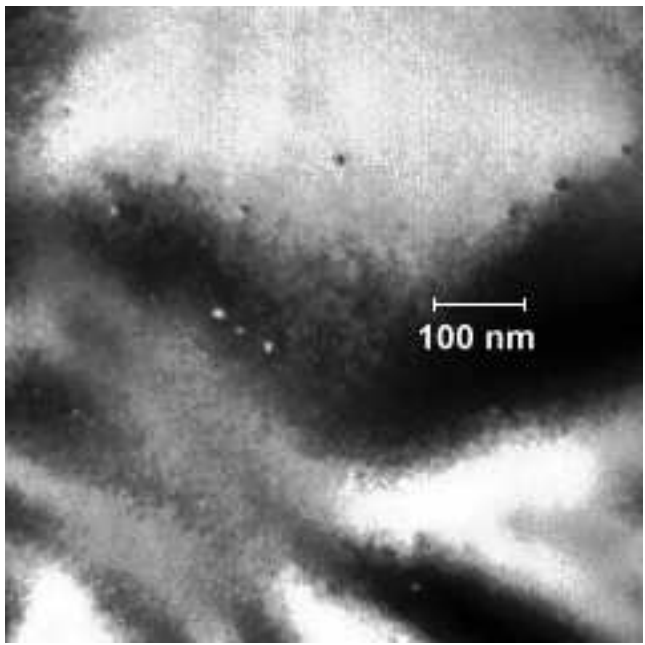}%
\caption{Defects which formed spontaneously when a (111) silicon specimen was given less than 5 minutes of air exposure after being thinned by Argon ion milling.  The greyvalue histogram of the image has been equalized to improve defect visibility.}
\label{Fig1}
\end{figure}

\begin{figure}[tbp]
\includegraphics{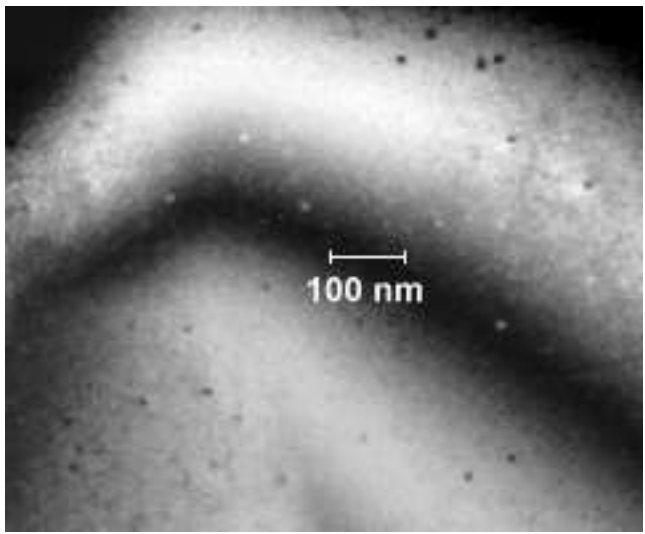}%
\caption{Defects seen with help from thickness fringe contrast under (220) 2-beam near-Bragg conditions in (001) silicon after a 3:1 nitric:HF acid etch.  Defect density and specimen thickness are not proportional.  The wedge half-angle is $\sim 12^{\circ}$.}
\label{Fig2}
\end{figure}

\begin{figure}[tbp]
\includegraphics{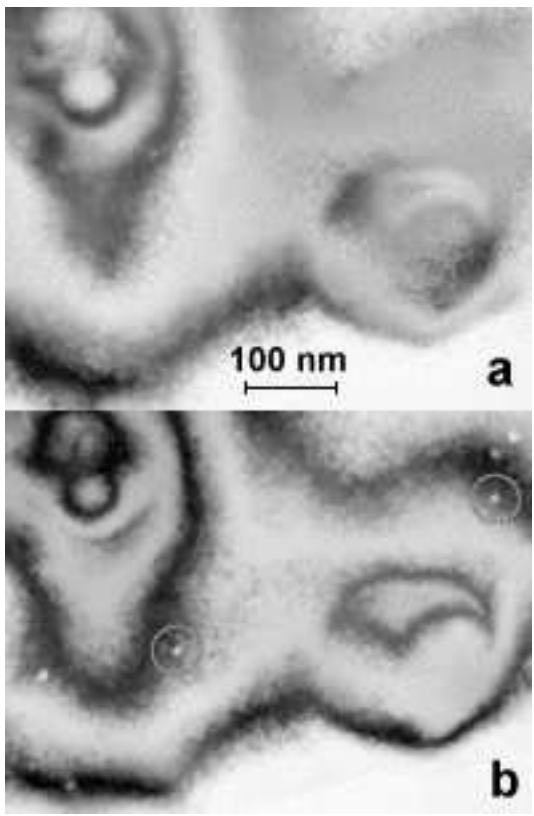}%
\caption{Images taken 1 and 45 hours after a 5 minute air exposure of $(100)$ silicon etched in 3:1 nitric:HF and then dipped in concentrated HF to retard the rate of oxidation.  We have detailed chronology on the formation time of all defects in the ``after" image, except for those circled.  These were formed in an interval between the 31 hour and the 45 hour mark, when no intervening observations were made.}
\label{Fig3}
\end{figure}

Figures \ref{Fig1}, \ref{Fig2}, and \ref{Fig3} show defects in specimens prepared with the above three methods.  All defects measured 5 to 15 nm in diameter, and had weak strain fields when, for example, compared with "bulk" oxygen precipitates of comparable size.  The defect density and the morphology of defects so detected are comparable in all specimens examined so far.  These include (100) specimens prepared by each of the methods mentioned above, plus ion milled specimens with (111) and (110) 
surface orientations.  

The defects are most easily seen under two-beam conditions in specimens with large wedge angles (Fig. \ref{Fig2}).  Here their contrast suggests that they're associated with ``beam-direction columns" in the specimen containing less silicon than found in adjacent columns, as expected for voids or amorphous precipitates.  The thickness of missing silicon (determined from the phase lag in contrast given the extinction distance for the active reflection) is comparable to the lateral size of the defects, suggesting that they are nearly as deep as they are wide.  Plots of defect number density of defects versus specimen thickness in several such images did not show the increase in projected density expected for defects distributed throughout the volume of the wafer.  The surface-correlated nature of these defects received further support from ion-milling experiments.  Removal of 10 to 30 nm of silicon (by 10 seconds of Argon ion milling) from a previously-examined surface created a new set of defects not correlated with those present before milling.  

\begin{figure}[tbp]
\includegraphics{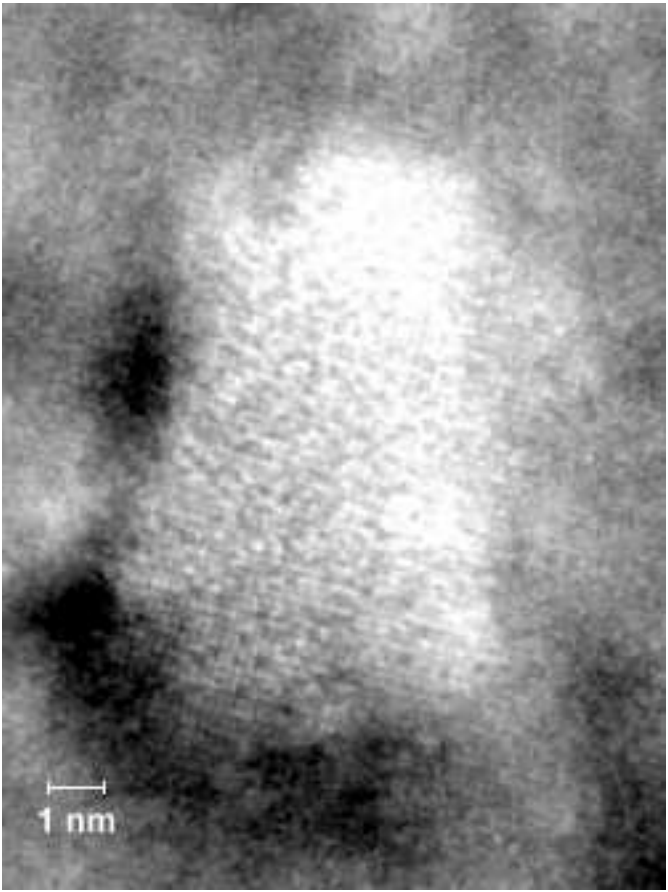}%
\caption{A HREM image of spontaneously-forming defects in an HF:nitric etched (100) silicon specimen.  Faceting parallel to the 0.192 nm (022) fringes is atypically strong in this defect.}
\label{Fig4}
\end{figure}

A high resolution TEM image of one defect is shown in Fig. \ref{Fig4}.  The defects show increased electron transmission on lateral size scales large compared to the transverse coherence width of the electrons \cite{Spence88} used in the experiment ($< 2 nm$).  They also show $(220)$ fringe contrast \cite{AllpressAndSanders} consistent with thinner silicon under the defect than adjacent to it.  On $\langle 100 \rangle$ surfaces, the defects sometimes show faceting along $(011)$, reminescent of (much larger) platelet oxygen precipitates in silicon \cite{TanTice76}.

Studies of formation kinetics have been attempted for $(100)$ specimens prepared by each of the specimen preparation protocols mentioned above.  Air exposure was minimized to between 1 and 10 minutes, following surface preparation by milling or wet chemistry, by rapidly inserting the specimen into the Ti getter-pump vacuum of the microscope.  Following a nitric-HF etch, this precaution resulted in no obvious decrease in the number of defects.  However, for ion-milled and HF-dipped specimens the number density of defects observed following insertion in the microscope was lower, by more than a factor of 10, than the density otherwise routinely observed (e.g. with post-preparation air exposures of half day or more).  

More importantly, on those specimens with reduced defect density, new defects began to appear in the microscope during ``time-lapse observation" of the specimens at 1 hour intervals over periods of days (Fig. \ref{Fig3}).  A preliminary quantitative look at the early stages of this process suggests e-folding times of approximately a day for our ion-milled specimen, and more like a half day for the etched surface after an HF-dip.  Re-exposure to air for a couple of hours, after a day or two in the microscope, results in a temporary increase in formation rate when the specimen is put back into the microscope.  However, the process of defect formation in the microscope seems to be ``already completed'' if the surfaces have instead been exposed to air for significantly longer periods of time (e.g. a day or more) before examination.

We have been unable to ``induce formation" of these defects by exposure to the electron beam, even at beam intensities orders of magnitude higher than those used for routine observation.   High resolution studies of these defects during the time series on HF-dipped specimens confirm our impression, from conventional TEM images, that the defects reach nearly their final size by first glance, i.e. on time scales short compared to an hour.  

Hence the formation of the defects seems, under these conditions, to be limited by rate of nucleation and not by rate of growth.  The initial rate of nucleation, in turn, may be limited by the surface concentration of a molecular species available through exposure to air (like oxygen).  Time-lapse series, and video tapes, over shorter time intervals should allow a look at the growth process itself.  Given the apparent insensitivity of the process to electron examination, such series may also allow lattice-scale "pre-nucleation" imaging of sites at which defects are destined to develop at a later time.

\section{Discussion}

The classic empirical test for artifacts of TEM specimen preparation is simple:  Thin the specimen by two physically different techniques, e.g. by both Ar ion-milling and by chemical etching.  Only defects that are not a result of specimen preparation are likely to be common in both cases.  This test was the inspiration for work on etched specimens, after we began characterizing more carefully the defects in Ar ion-milled silicon mentioned above.  Even though the literature \cite{Bangert86} argued against Argon bubbles as a possible source, given our high incidence angles and relatively large defect sizes, we wanted experimental confirmation.  

The defects pass the classical test, in that they are ubiquitous in etched and ion-milled surfaces of many orientations and vintage (some stored in air for $> 2$ decades).  Work here, however, shows that they do form after specimen preparation.  In that sense, literature studies of bulk defects have appropriately ignored them.  The presence of comparable defect number densities, sizes, and in the case of ion milled and HF-dipped specimens comparable formation dynamics, after physically diverse thinning processes thus argues that they result from the interaction between fresh silicon surfaces and air or the microscope vacuum, independent of the way in which the surfaces are created.  

Observed parameters are as follows:  (i) Final defect number densities, after extended air exposure at standard temperature and pressure are $\sim 10^{10}$ defects/$cm^2$.  (ii) Initial e-folding times for nucleation, independent of air exposure, are in the 10 to 30 hour range.  (iii) Final size of defects is $\sim 10 nm$, sometimes trailing to smaller and less visible sizes.  (iv) Time elapsed between nucleation, and growth to full size, is much less than an hour.  

Why these values?  Note that $10^{10}$ defects/$cm^2$ corresponds to an average half-distance between defects of around 50 nm.  This is comparable to the distance for strain field relaxation in silicon \cite{AshbyBrown63}.  The defects might thus support a process of silicon surface ``hardening" on air exposure, in which a compressional strain field (still mild compared to that of oxygen precipitates in a wafer interior) is built up around an array of defects due to oxidation.  Oxidation of silicon surfaces occurs at room temperature by indiffusion that can leave atomic steps on the surface statistically intact.  Hence significant expansion likely takes place at the buried oxide/silicon interface.  The resulting strain field, much like the reaction ``skin" on a pot of soup allowed to sit, could in turn limit further defect formation and defect size.  One consequence of this mechanism would be an expected decrease in size of intrusions nearby one another, something not demonstated so far.

\begin{figure}[tbp]
\includegraphics{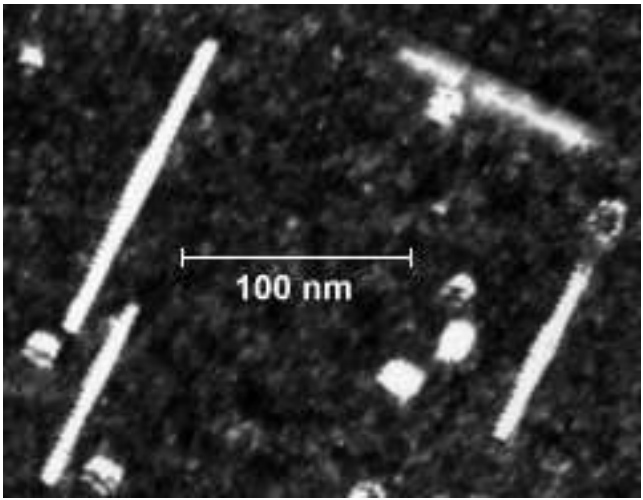}%
\caption{Image of crystalline defects with comparably weak strain fields, small sizes, and high number densities likely formed after thinning (100) silicon.}
\label{Fig8}
\end{figure}

The presence of such subsurface defects will have consequences.  Even buried beneath the native oxide, they will scatter light diffusely.  Harada et al (\cite{Harada00}) report SPM observation of hillocks on (100) Si that assist early stage oxidation, with lateral sizes and number densities comparable to the defects reported here.  Silicon's volume increase on oxidation may thus result in associated surface hillocks.  This is supported so far by preliminary work in our lab at creating such defects on extremely flat (as evidenced by 0.13 nm steps after native oxidation) epitaxial silicon.  We now think that previously reported crystalline defects in TEM specimens of ``bulk'' silicon (Fig. \ref{Fig8}) with comparable number densities and a $2\%$ lattice misfit \cite{Fraundorf89} may also have formed at specimen surfaces after thinning.  Recent HREM work on similarities between the above crystalline defects and $Cu_3 Si$ colony defects \cite{Solberg78} in bulk silicon, and the similarity of intrusions reported here to S-pit ``intrusions'' associated with nickel contamination \cite{Gail83} in Si, suggest that heterogeneous nucleation may play a role in determining the number density of these defects as well.

\begin{acknowledgments}

  Thanks to Lu Fei, Jeff Libbert, Lucio Mulestagno, and Blake Rowe at MEMC Electronic Materials for insight and diverse specimens, and to MEMC, Monsanto, and Boeing for regional facility support. 

\end{acknowledgments}

\bibliography{osirefs1}

\end{document}